\documentclass[conference,10pt]{IEEEtran}
\usepackage[shortcuts,acronym]{glossaries}

\newacronym{ml}{ML}{Machine Learning}
\newacronym{dl}{DL}{Deep Learning}

\newacronym{gnn}{GNN}{Graph Neural Network}
\newacronym{gcn}{GCN}{Graph Convolutional Network}
\newacronym{fcn}{FCN}{Fully Connected Network}

\newacronym{fmcw}{FMCW}{Frequency-Modulated Continuous-Wave}
\newacronym{mti}{MTI}{Moving Target Indication}
\newacronym{fft}{FFT}{Fast Fourier Transform}
\newacronym{rdi}{RDI}{Range Doppler Image}
\newacronym{rai}{RAI}{Range Angle Image}
\newacronym{cnn}{CNN}{Convolutional Neural Network}
\newacronym{dgcnn}{DGCNN}{Dynamic Graph CNN}
\newacronym{doa}{DOA}{Direction Of Arrival}
\newacronym{mse}{MSE}{Mean Squared Error}
\newacronym{lstm}{LSTM}{Long short-term memory}
\newacronym{rcnn}{R-CNN}{Regions with \ac{cnn} features}
\newacronym{mlp}{MLP}{Multi-Layer Perceptron}

\usepackage{lipsum}
\usepackage{todonotes}
\usepackage[english]{babel}               
\usepackage{amsmath}
\usepackage{amsfonts}
\usepackage{graphicx}
\usepackage{subcaption}
\usepackage[font=small,labelfont=bf]{caption} 
\usepackage{booktabs}
\usepackage{multirow}
\usepackage{array}
\usepackage{algorithmic}
\usepackage{algorithm}
\usepackage{hyperref}
\usepackage{tikz}
\usepackage{pgfplots}
\usepgfplotslibrary{groupplots}
\usepackage{float}
\usepackage{epstopdf}
\usepackage{enumitem}
\usepackage{setspace}
\usepackage{nicefrac}
\usepackage{url}

\usepackage{bm}
\usepackage{mathtools, nccmath}
\usepackage{svg}
\usepackage{pdfpages}
\pgfplotsset{compat=1.17}
\begin{document}
\IEEEpubid{\makebox[\columnwidth]{0000--0000/00\$00.00 $\copyright$2015 IEEE \hfill} \hspace{\columnsep}\makebox[\columnwidth]{ }}
\title{Cross-modal Learning of Graph Representations using Radar Point Cloud for Long-Range Gesture Recognition}
% \IEEEspecialpapernotice{(Invited Paper)}
%\author{\IEEEauthorblockN{{Souvik Hazra$^{1*}$, Hao Feng$^{1*}$, Gamze Naz Kiprit$^{1}$, Michael Stephan$^{1,2}$,Lorenzo Servadei$^{1,3}$, Avik Santra$^{1}$, Robert Wille$^{3}$,\\ Robert Weigel$^{2}$}}
%\IEEEauthorblockA{\textit{$^{1}$Infineon Technologies AG}, Neubiberg, Germany\\
%$^{2}$Friedrich-Alexander-University Erlangen-Nuremberg, Erlangen, Germany\\
%$^{3}$Technical University of Munich, Munich, Germany\\
%E-mail: \{souvik.hazra,lorenzo.servadei, avik.santra\}@infineon.com\\
%\{robert.wille\}@tum.de\\
%\{michael.stephan, robert.weigel\}@fau.de\\
%$^*$equal contribution}}

\author{\IEEEauthorblockN{Souvik Hazra\IEEEauthorrefmark{1}\IEEEauthorrefmark{4}, Hao Feng\IEEEauthorrefmark{1}\IEEEauthorrefmark{3}\IEEEauthorrefmark{4}, Gamze Naz Kiprit\IEEEauthorrefmark{1}\IEEEauthorrefmark{3}, Michael Stephan\IEEEauthorrefmark{1}\IEEEauthorrefmark{2}, Lorenzo Servadei\IEEEauthorrefmark{1}\IEEEauthorrefmark{3}, Robert Wille\IEEEauthorrefmark{3}, \\ Robert Weigel\IEEEauthorrefmark{2}, Avik Santra\IEEEauthorrefmark{1}}
\IEEEauthorblockA{\IEEEauthorrefmark{1}Infineon Technologies AG, Neubiberg, Germany}
\IEEEauthorblockA{\IEEEauthorrefmark{2}Friedrich-Alexander-University Erlangen-Nuremberg, Erlangen, Germany}
\IEEEauthorblockA{\IEEEauthorrefmark{3}Technical University of Munich, Munich, Germany} 
\IEEEauthorblockA{Email: \{souvik.hazra, lorenzo.servadei, avik.santra\}@infineon.com} \IEEEauthorblockA{Email: \{robert.wille\}@tum.de}
\IEEEauthorblockA{E-mail: \{michael.stephan, robert.weigel\}@fau.de}
\IEEEauthorblockA{\IEEEauthorrefmark{4}equal contribution}}

\maketitle
\begin{abstract}
Gesture recognition is one of the most intuitive ways of interaction and has gathered particular attention for human computer interaction. Radar sensors possess multiple intrinsic properties, such as their ability to work in low illumination, harsh weather conditions, and being low-cost and compact, making them highly preferable for a gesture recognition solution. However, most literature work focuses on solutions with a limited range that is lower than a meter. We propose a novel architecture for a long-range (1m - 2m) gesture recognition solution that leverages a point cloud-based cross-learning approach from camera point cloud to 60-GHz FMCW radar point cloud, which allows learning better representations while suppressing noise. We use a variant of \ac{dgcnn} for the cross-learning, enabling us to model relationships between the points at a local and global level and to model the temporal dynamics a Bidirectional \ac{lstm} network is employed. In the experimental results section, we demonstrate our model's overall accuracy of 98.4\% for five gestures and its generalization capability.
\end{abstract}
\section{Introduction}
% Souvik Will do it 
Human control interfaces for various indoor applications, e.g., sound systems, television, and lightning, continuously evolve to increase user comfort.
Going from buttons, switches, and rotary knobs, to touchscreen-based control with smartphone-apps, the trend is now heading towards gesture-based control \cite{gesture_smart_home_radar, ges_radar_car}, allowing for interaction from further distances without requiring additional hardware.
Gesture control needs to be intuitive, accurate, and should be computationally cheap \cite{gesture_recog_4154947}. While vision-based systems nowadays fulfill these criteria, they are often seen as too privacy-invasive. 
Therefore, radar-based systems have recently been under investigation as a privacy-preserving alternative solution \cite{Hayashi2021,Stephan2021a,Stephan2022,servadei2021label,santra2020deep}.
However, while progress has been made, and radar-based gesture sensing has been deployed in products \cite{Hayashi2021,8542778,8999087,7131232},
it is still less accurate than vision-based systems, especially on longer distances.
Therefore, we present a mmWave radar-based gesture-recognition solution that incorporates knowledge from a camera system during the training process.
Our proposed system uses preprocessed radar data, in the form of point clouds, as input to a \ac{dgcnn} \cite{wang2019dynamic} for performing cross-learning from a camera point cloud, which then predicts the performed gesture, taking multiple frames into account.

% Could add description of paper content here, but probably wastes too much space
%\section{Background}
% \input{chapters/background}
\section{Camera Point Cloud}
\begin{figure}
    \begin{subfigure}[b]{.5\columnwidth}
        \includegraphics[width=\linewidth, trim={8mm 8mm 8mm 8mm},clip]{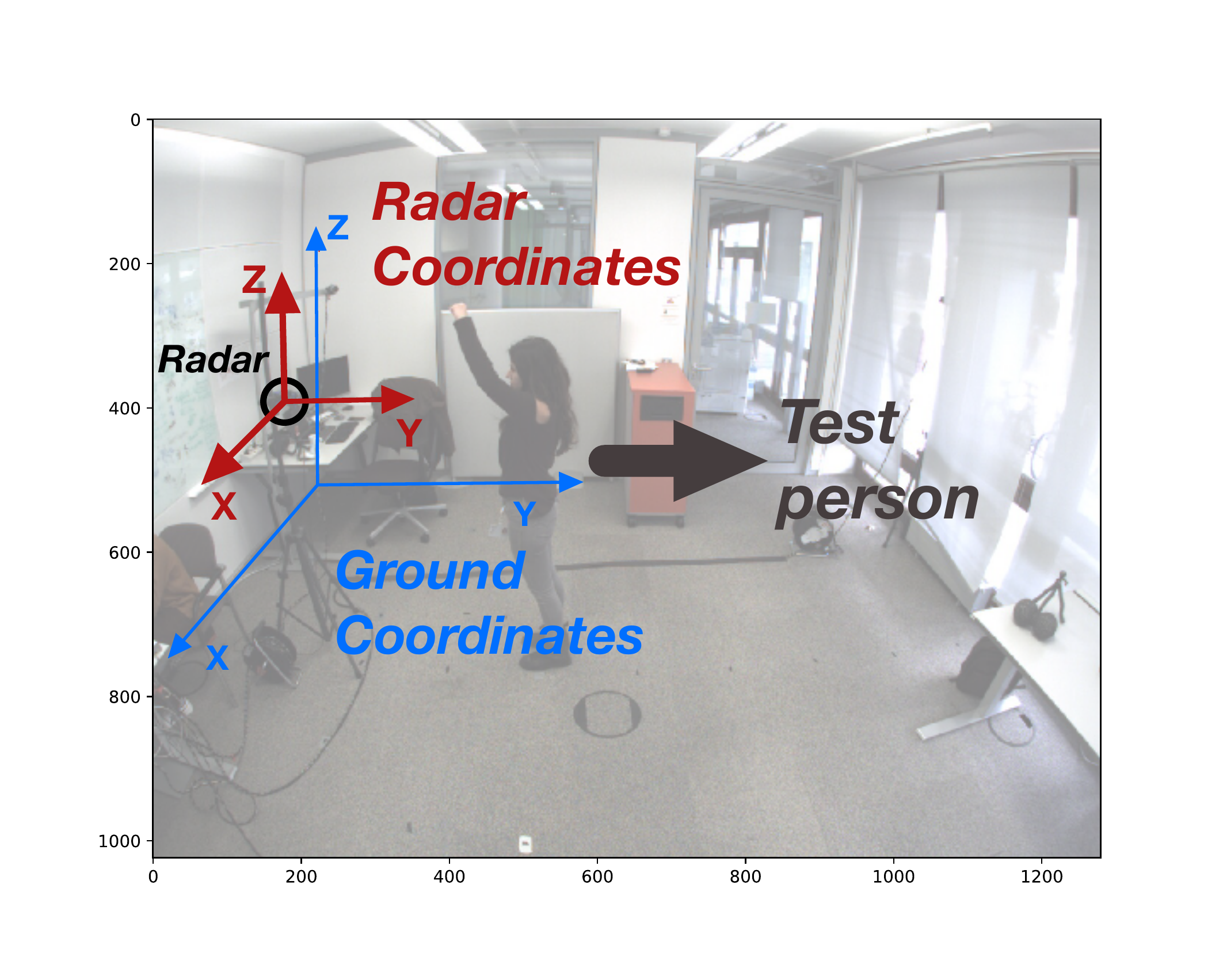}
        \caption{Radar setup. $\theta_{tilt} = 0$. }
        \label{fig:radar_setup}
    \end{subfigure}
    \hfill
    \begin{subfigure}[b]{0.49\columnwidth}
        \includegraphics[width=\linewidth, trim={20mm 10mm 18mm 20mm},clip]{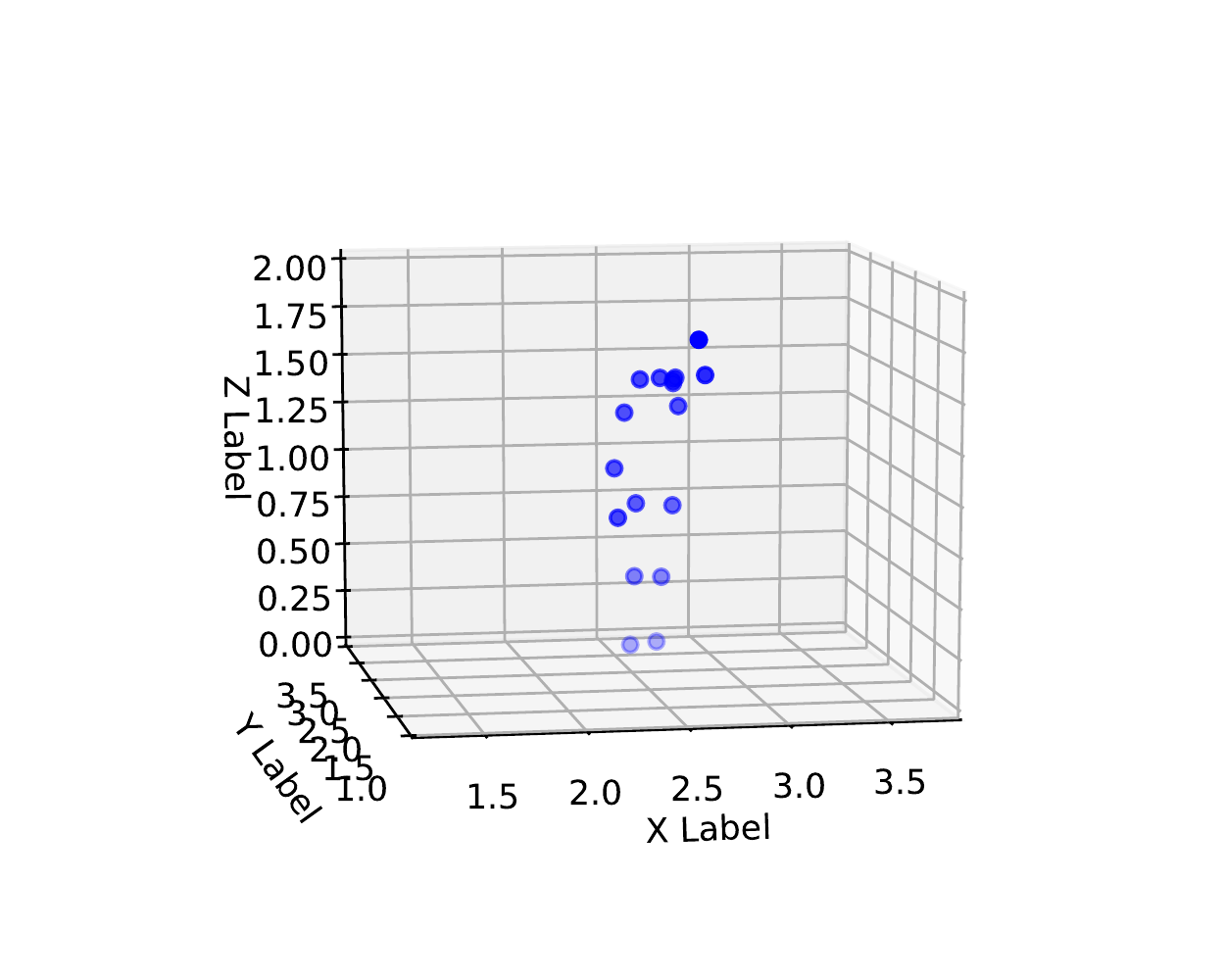}
        \caption{Camera point cloud.}
        \label{fig:skeletons}
    \end{subfigure}
    \hfill
    \begin{subfigure}[b]{0.49\columnwidth}
        \includegraphics[width=\linewidth, trim={2mm 0mm 10mm 18mm},clip]{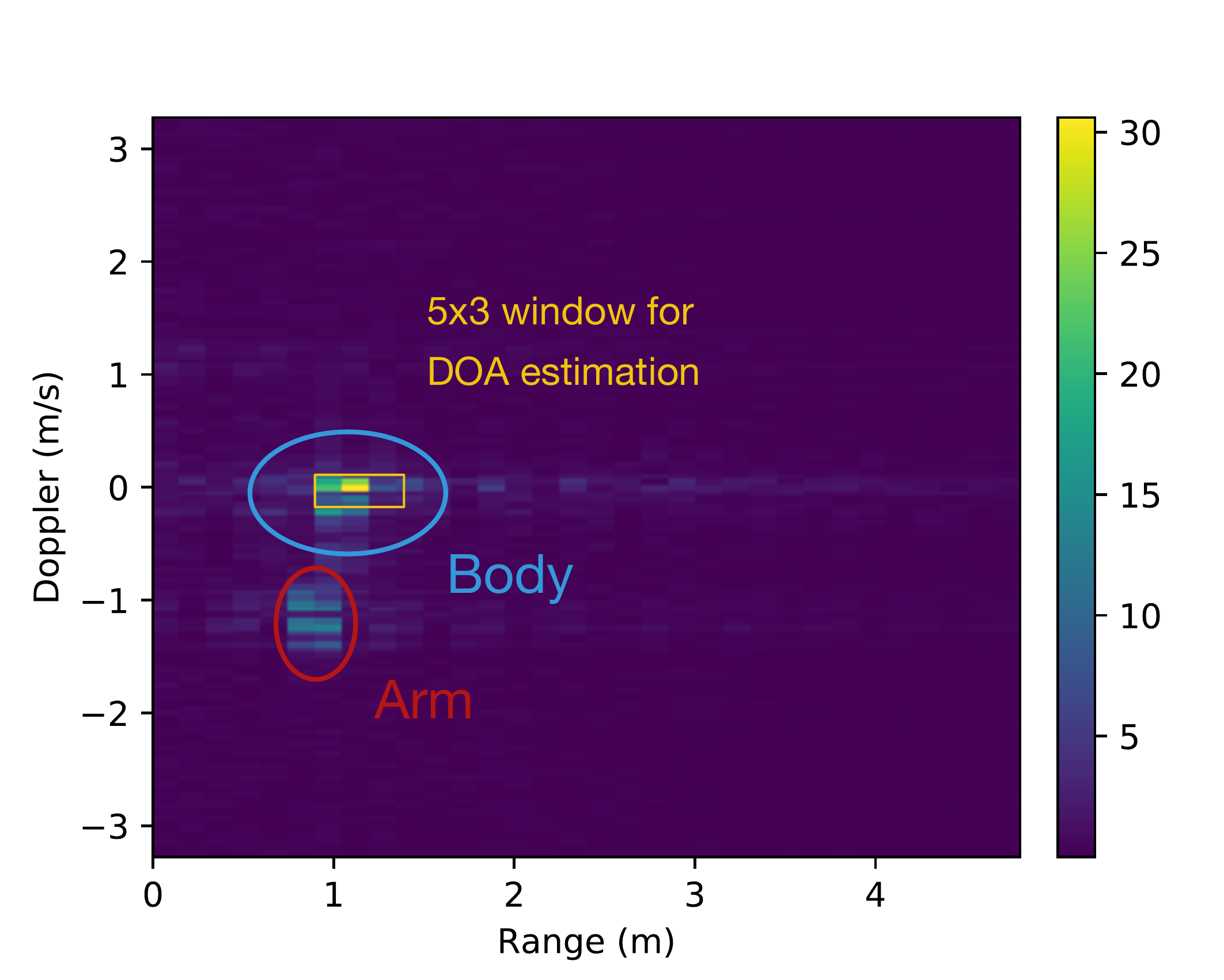}
        \caption{Range-Doppler Image. Color bar shows the intensity of each bin. }
        \label{fig:rdi}
    \end{subfigure}
    \hfill
    \begin{subfigure}[b]{0.5\columnwidth}
        \includegraphics[width=\linewidth, trim={28mm 8mm 8mm 2mm},clip]{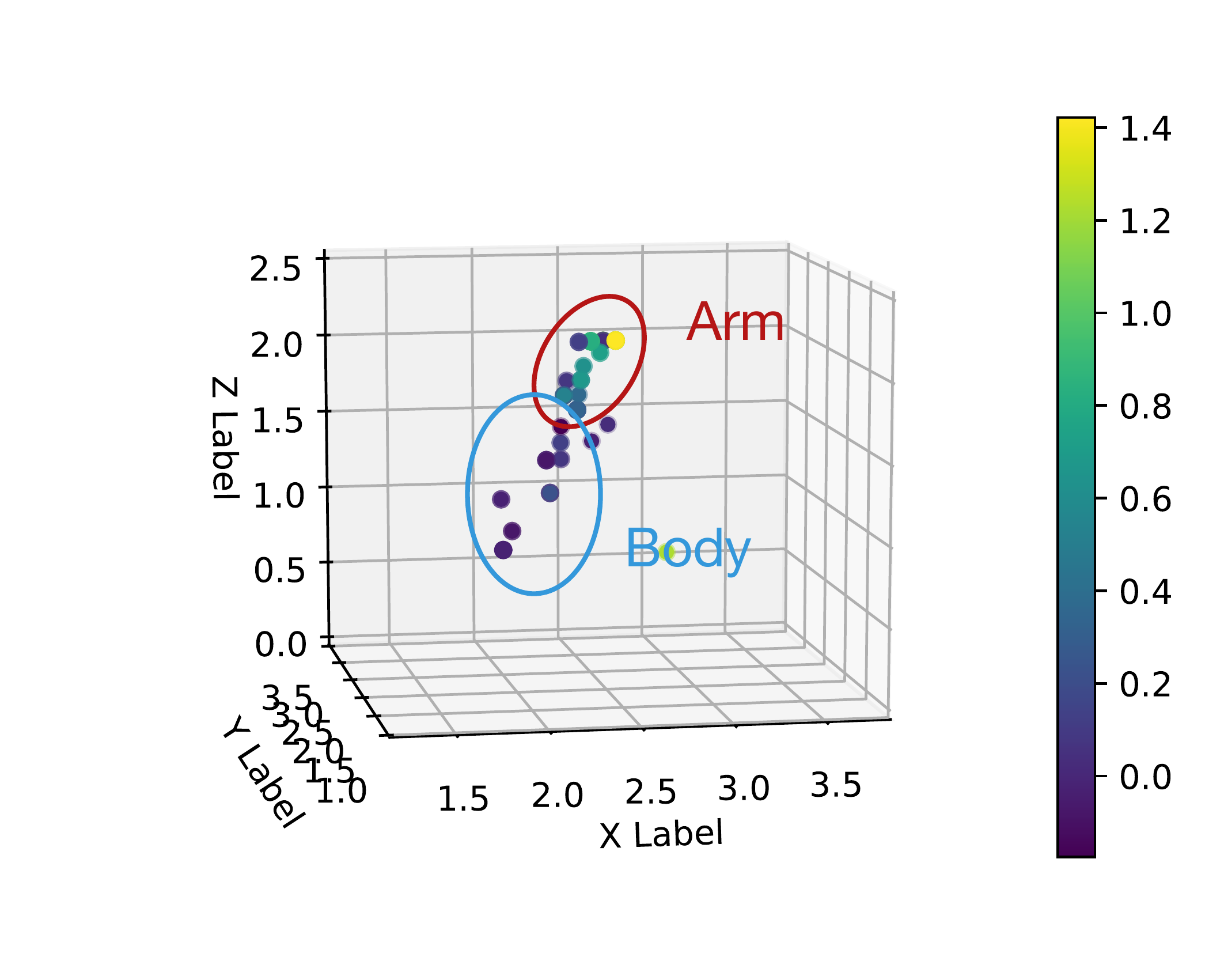}
        \caption{Radar point cloud with color showing velocity value in m/s. }
        \label{fig:radar_pc}
    \end{subfigure}
\caption{Visualization during anti-clockwise gesture. }
\label{fig:visualization}
\vspace{-2mm}
\end{figure}

\begin{figure*}[t!]
    	\begin{center}
    	    \vspace{-0mm}
    		\includegraphics[width=0.9\textwidth]{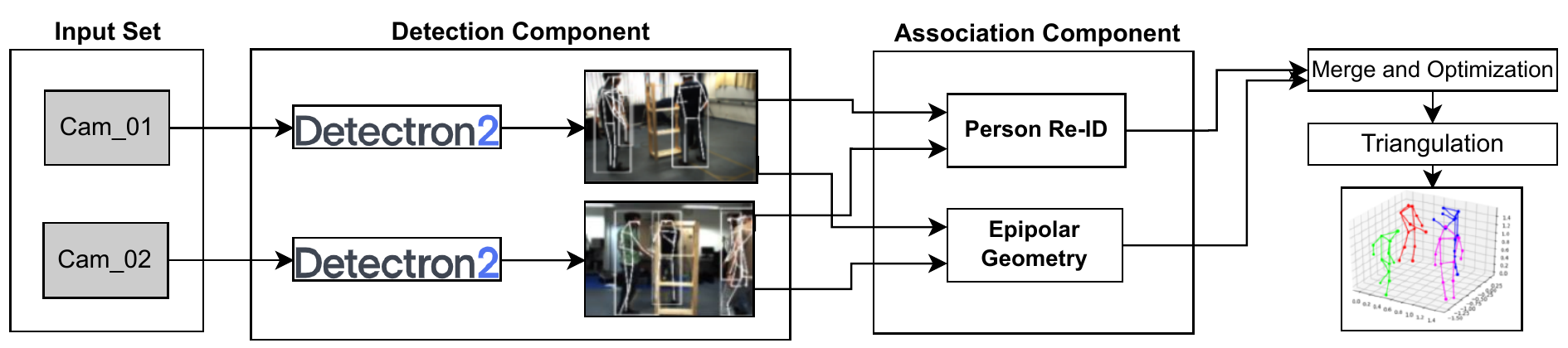}
    		\vspace{-2mm}
    	\end{center}
    	\caption{\textbf{Overview of used approach for skeleton detection.} Given an input set of calibrated camera recordings from different views, we first detect persons and their keypoints in 2D space using Detectron2. Thus, to match 2D poses in multiple views, we combine person re-identification and geometric information of persons. The person re-identification cues along with the geometric cues are then concatenated and triangularized to determine 3D points.}
    	\vspace{-2mm}
    	\label{fig:skeletondetectionposter}
\end{figure*}

%This section briefly reviews the method used to obtain camera point clouds and associated background concepts.
%\subsection{Skeleton Detection}
To determine 3D joint coordinates from 2D camera images, the joints must be visible in at least two cameras and must be matched. First, we feed all camera images into Detectron2 \cite{wu2019detectron2} to detect people and their associated keypoints in 2D space. Here we use the best pretrained keypoint \ac{rcnn} architecture available in the Detectron2 model zoo trained on the Coco keypoint dataset \cite{cocodataset}. The Coco keypoint dataset contains 17 keypoints for each person, which are the nose, left eye, right eye, left ear, right ear, left shoulder, right shoulder, left elbow, right elbow, left wrist, right wrist, left hip, right hip, left knee, right knee, left ankle, right ankle. Matching 2D poses through multiple views is challenging for various reasons, such as occlusion and truncation. Our association component, as shown in Fig. \ref{fig:skeletondetectionposter}, consists of two robust approaches that complement and reinforce each other: person re-identification and epipolar geometry. The result of our approach is as shown in Fig. \ref{fig:skeletons}, 3D joint coordinates, and 17 keypoints.
\subsubsection{Person Re-Identification}
The goal of person re-identification is to identify the corresponding person of interest in different views or at different times. We use the implementation provided in \cite{personreid}. It uses a \ac{ml} model, in which the embeddings of individuals that are close to each other are considered to be the same person.
\subsubsection{Epipolar Geometry}
%When multiple cameras view the same scene with different views, it is not surprising that some geometric information comes out. 
Using epipolar geometry, the geometric information can be leveraged and objects localized in different cameras. For example, if the object's location is specified in the first camera, the search area in the second camera is constrained to a single line if the epipolar geometry is known \cite{epipolargeometry}.

%%%%%%%Overview figure of used approach%%%%%

\section{Radar Point Cloud}
\begin{figure}[H]
    \centering
	\includegraphics[width=\columnwidth]{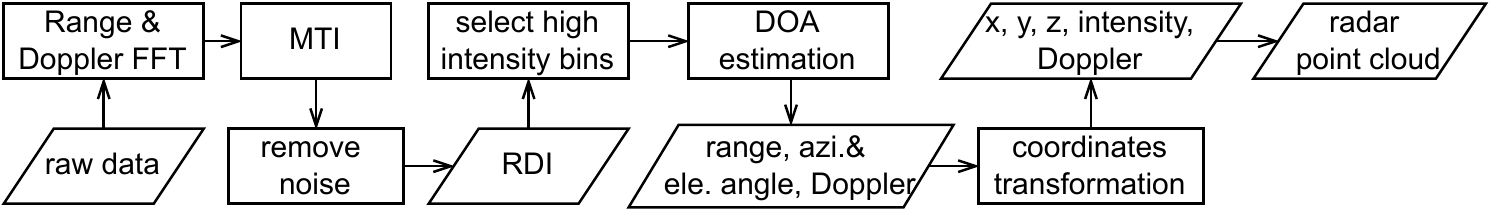}
	\caption{Pre-processing of raw radar data.}
	\label{fig:preprocessing}
\end{figure}

This section describes the radar sensor setup and the preprocessing for converting raw radar data to 5-dimensional point clouds \cite{tesla_rapture} with x-y-z coordinates, intensity, and Doppler values, as illustrated in Fig. \ref{fig:preprocessing}.

%%%%%%%Flowchart of Pre-processing%%%%

\subsection{mmWave \acrshort{fmcw} Radar Sensor}
%\begin{figure}[H]
%    \centering
%    \includegraphics[width=0.3\linewidth]{figures/radar_chipset_coin.jpeg}
%    \caption{\emph{Infineon}'s \emph{BGT60TR13C} 60-GHz radar sensor.}
%    \label{fig:radar}
%    \vspace{-0.2cm}
%\end{figure}
We use Infineon's BGT60TR13C FMCW radar chipset. It operates by transmitting multiple frames, each containing a sequence of frequency chirps with a short ramp time.
%and delays between them which makes it the most commonly used as it saves power and computation requirement for post-processing.
The response is digitized in 12-bit by the analog-digital converter, and is further passed to the PC over USB. The operating parameters of the radar are presented in Tab. \ref{System_Parameters}.
%\vspace{0.6cm}
 {\renewcommand{\arraystretch}{0.8}
	\begin{table}[H]
		%\caption{Operating Parameters.} 
		\centering
		\setlength{\tabcolsep}{15pt}
		\begin{tabular}{ c|c  } 
			\hline
			\textbf{Parameters, Symbol}  & \textbf{Value }   \\
			\hline
			\hline
			Ramp start frequency, $f_{\text{min}}$  & 57.5 GHz    \\
			%\hline
			Ramp stop frequency , $f_{\text{max}}$ & 58.5 GHz    \\
			%\hline
			
		    Frame rate, $f$ & 20 fps    \\
			%\hline
		
			Number of samples per chirp, $\text{NTS}$ & 64  \\
			%\hline
		
			Sampling frequency, $fs $&   2 MHz  \\
			%\hline
			Chirp time, $T_c $   & 64 $\mu$s\\
			%\hline
		
			Number of chirps, $\text{PN} $&   128  \\
			%\hline

			Number of Tx antennas, $N_{\text{Tx}}$    & 1  \\
			%\hline
			Number of Rx antennas, $N_{\text{Rx}}$    & 3  \\
			%\hline
			
			\hline
		\end{tabular} 
		\caption{Operating Parameters.} 
		\label{System_Parameters}
	\end{table}
%	\vspace{-0.3cm}
}
%\vspace{-0.5cm}
\subsection{Point Cloud Transformation}
\subsubsection{\ac{rdi} and Angles Estimation}
The reflected target signal is mixed with the transmitted chirp signal and then passed through a low-pass filter to obtain the intermediate frequency signal \cite{ldc_ges_fmcw}. Since we are only interested in moving targets in gesture recognition, we use previous frame subtraction as \ac{mti} \cite{comprehensive_mpoint}.
% By applying 1D \acrfull{fft} to the sampled IF signal along fast time, %which we refer to as Range-\acrshort{fft}
% , and another 1D \acrshort{fft} on the slow time axis, referred to as Doppler-\acrshort{fft}, we can acquire \acrshort{rdi} comprising range, doppler and intensity information.
By applying 2D Range-Doppler-\ac{fft} along the sample and chirp axes, we can acquire \acp{rdi}, comprising range, Doppler, and intensity information, as shown in Fig. \ref{fig:rdi}. 
Additionally, as we work with single-person scenarios, we first detect the range-Doppler bin with the highest intensity and filter out any points far away from it as noise \cite{radar_pointgnn}. Afterwards, we select a certain number of high-intensity range-Doppler bins and use the values in a $5\times3$ window around each bin, seen in Fig \ref{fig:rdi}, to estimate the covariance matrix for \ac{doa} estimation with bartlett beamforming.
Now the points are in spherical radar coordinates $(r,  \theta_{azi},  \theta_{ele})$, which correspond to range, azimuth angle, and elevation angle. We finally apply the transformation matrix \cite{jin2020mmfall} shown below,

\begin{equation}
\resizebox{.8\hsize}{!}{$
\begin{bmatrix}
x \\y \\z
\end{bmatrix} =
\begin{bmatrix}
1 & 0 & 0 \\
0 & \cos{\theta_{tilt}} & \sin{\theta_{tilt}} \\
0 & -\sin{\theta_{tilt}} & \cos{\theta_{tilt}}
\end{bmatrix} 
\begin{bmatrix}
r\cos{\theta_{ele}}\sin{\theta_{azi}} \\
r\cos{\theta_{ele}}\cos{\theta{azi}} \\
r\sin{\theta_{ele}}
\end{bmatrix} + 
\begin{bmatrix}
x_{r} \\ y_{r} \\ h
\end{bmatrix}$}
\end{equation}

where $(x,y,z)$ denote the resulting ground Cartesian coordinates, $\theta_{tilt}$  and $ (x_{r}, y_{r}, h) $ are the tilt angle and ground Cartesian coordinates of radar, respectively. 

\subsubsection{Point Cloud}
With the help of a radar sensor, we can collect the points corresponding to the detected object. However, a specific number of sample points for each frame is required as input. Thus, to simplify the computation, traditional zero-padding is the method we choose for oversampling, which means adding all-zero points to complement the point cloud. Meanwhile, for frames with more points than we require, sorting the points by their intensity and selecting a predefined number allows us to save the samples with the most information \cite{3d_radar_pointcloud_gen}. Finally, we take $(x, y, z, d, intensity)$ which are coordinates, Doppler, and intensity value into consideration as the features of each point, shown in Fig. \ref{fig:radar_pc}. 

%%%%%%%Radar Point Cloud%%%%
% \begin{figure}[H]
%     \centering
% 	\includegraphics[width=0.50\textwidth]{figures/radar_pointcloud_marked.png}
% 	\caption{Radar point cloud.}
% 	\label{fig:radar_pointcloud}
% \end{figure}

%%%%%%%Figure of radar setup%%%%
%\begin{figure}[H]
%    \centering
%	\includegraphics[width=0.40\textwidth]{figures/radar_setup_ma%rked.jpg}
%	\caption{Radar set-up.}
%	\label{fig:radar_setup}
%\end{figure}

\section{Architecture and Learning}
\begin{figure*}
    	\begin{center}
    	    \vspace{-0mm}
    	\includegraphics[width=0.9\textwidth]{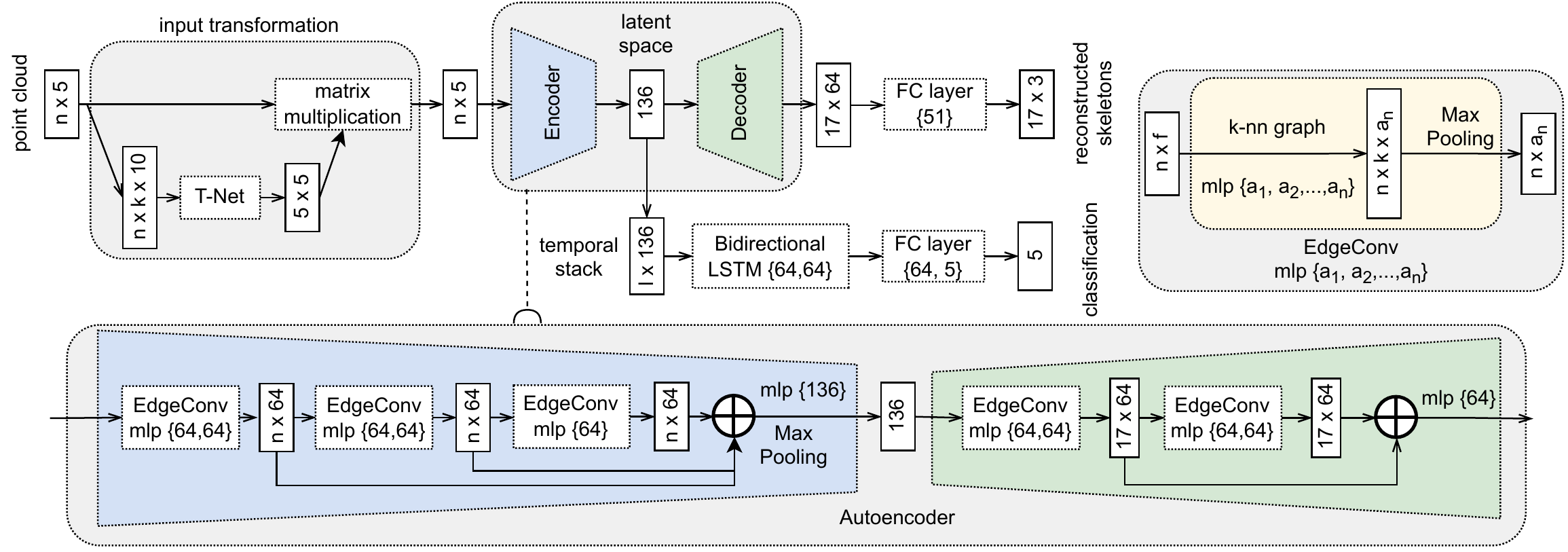}
    		\vspace{-1mm}
    	\end{center}
    	\caption{\textbf{Network Architecture:} Radar point cloud in shape $n\times5$ first passes through an input transformation module and then is fed into \ac{dgcnn} Autoencoder, followed by one fully connected layer to reconstruct camera-based skeletons. Next, the latent space of each Autoencoder-frame is stacked in temporal order with sequence length $l=30$, which is used to generate classification scores for five classes by two bidirectional \ac{lstm} layers and two fully connected layers. \textbf{EdgeConv block:} A tensor of shape $n \times f$ is fed into the block as input, for computing edge features of each point in $k$-nn Graph by a \ac{mlp} with the number of layer units denoted as $\{a_1, a_2, \dots, a_n \}$. Then, after max pooling across neighboring edge features, we get a updated tensor of shape $n \times a_n$.}
    	\vspace{-6mm}
    	\label{fig:model_pipeline}
\end{figure*}

In this section, we introduce the network's architecture of our proposed Autoencoder shown in Fig. \ref{fig:model_pipeline} for cross-learning between radar and camera information \cite{rodnet}. % by applying \acrfull{dgcnn} \cite{wang2019dynamic}, followed by LSTM layers for recognition of gesture sequences
The architecture is composed of a \ac{gnn} \cite{9531070}, specifically \ac{dgcnn} \cite{wang2019dynamic}, followed by \ac{lstm} layers for recognition of gesture sequences.
%%%%%%%Network Architecture%%%%
\subsection{Edge Convolution} \label{section:edgeconv}
\begin{figure}[H]
    \centering
	\includegraphics[width=0.40\textwidth]{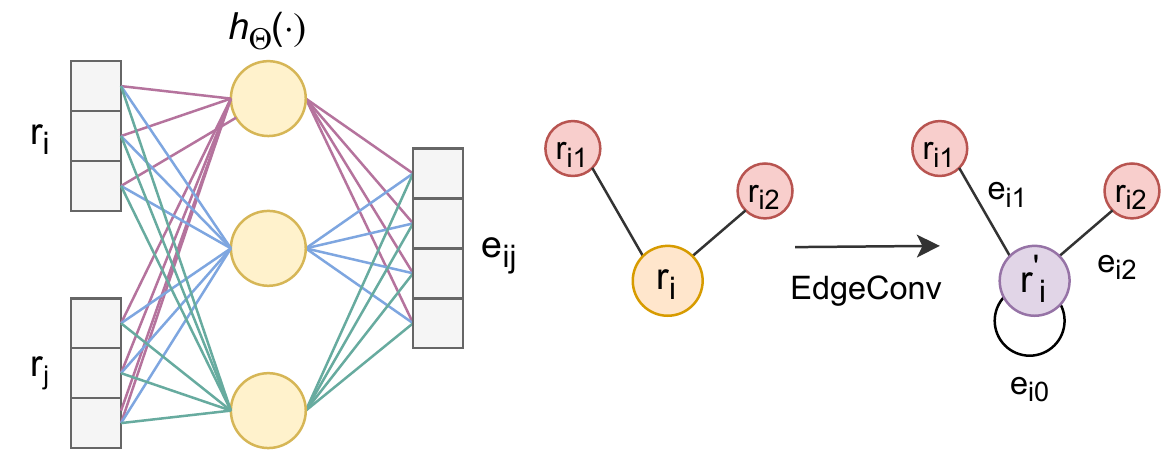}
	\caption{\textbf{Left}: Generating $f^{'}$-dimensional edge features $\mathbf{e}_{ij}$ from two $f$-dimensional points $\mathbf{r}_i$ and $\mathbf{r}_j$. Here, we choose one fully connected layer as $h_{\Theta} (\cdot)$ and $f=3$, $f^{'}=4$ for example. \textbf{Right}: Updating $\mathbf{r}_{i}^{'}$ from $\mathbf{r}_i$ by Edge Convolution, $k=3$ in our case, and $e_{i0}$ indicates self loop.}
	\label{fig:edgeconv}
\end{figure}

The radar point cloud contains $n$ points with $F$ features in each frame, denoted by $\mathrm{\mathbf{R} = \{\mathbf{r}_1,...,\mathbf{r}_n\}} \subseteq \mathbb{R}^F$. In our case, $F = 5$, representing 3D coordinates $x_i,y_i,z_i$, intensity and Doppler value $i,d$ in \acrshort{rdi} and $n$ points per frame. We apply $k$-Nearest Neighbour ($k$-NN) by Euclidean distance to generate the local graph of $\mathrm{\mathbf{R}}$ in $\mathbb{R}^F$ including self-loop, represented as $\mathcal{G} = (\mathcal{V},\mathcal{E})$, where $\mathcal{V} = \{1,...,n\}$ and $\mathcal{E} \subseteq \mathcal{V} \times \mathcal{V}$ are the $vertices$ and $edges$, respectively. As illustrated in Fig. \ref{fig:edgeconv}, to capture both global structure and local neighbourhood information, $edge features$ between two points is computed as 
\begin{equation}
    \bm{e}_{ij} =
     h_\mathbf{\Theta}(\mathbf{r}_i, \mathbf{r}_j-\mathbf{r}_i) 
\end{equation}
where $h_\mathbf{\Theta}: \mathbb{R}^F \times \mathbb{R}^F \rightarrow \mathbb{R}^{F^\prime}$ is a nonlinear function with learnable parameters $\mathbf{\Theta}$ and $\{\mathbf{r}_j: (i,j) \in \mathcal{E}\}$ is the set of neighbours around $\mathbf{r}_i$. More specifically, this edge convolution can be expressed as 
\begin{equation}
    %  e_{ijf^{\prime}} = \textrm{Leaky\_ReLU}(\bm{\theta}_{f^\prime}  \cdot(\mathbf{r}_j-\mathbf{r}_i)+\bm{\phi}_{f^\prime} \cdot \mathbf{r}_i),
     e_{ijf^{\prime}} = \sigma(\bm{\theta}_{f^\prime}  \cdot(\mathbf{r}_j-\mathbf{r}_i)+\bm{\phi}_{f^\prime} \cdot \mathbf{r}_i),
\end{equation}
and implemented as a shared \ac{mlp}, here 
$\mathbf{\Theta}=(\bm{\theta}_1,\dots,\bm{\theta}_{F^\prime},  \bm{\phi}_1,\dots,\bm{\phi}_{F^\prime})$, $leaky ReLU$ is chosen as non-linear function $\sigma(\cdot)$. At last, we take max pooling as the aggregation operation on the $edge features$ to update the points
\begin{equation}
    r^\prime_{if^\prime} = 
    \max_{j:(i,j)\in\mathcal{E}} e_{ijf^{\prime}}.
\end{equation}
which can capture the sharpest features to represent the points in lower-level.
In general, this EdgeConv creates an $F^\prime$-dimensional point cloud with the same number of points as the input $F$-dimensional point cloud, where $k=3$ and $n=64$. 

%%%%%%%Illustration of EdgeConv%%%%

\subsection{Model Architecture}
The overview of our whole network architecture is visualized in Fig. \ref{fig:model_pipeline}, including three main parts: input transformation, cross learning between radar and camera point cloud, and gesture recognition. 

\subsubsection{Input transformation}
As described in Section 3, we multiply the coordinates in radar coordination by our radar setup's rotation matrix to convert them to ground Cartesian coordinates. However, this transformation is insufficiently precise due to the measurement, and the object, in this case the test person, should be invariant following certain geometric modifications. As a result, we refer to the spatial transformation in this contribution \cite{DBLP:journals/corr/QiSMG16}, in which a mini-network (T-net in Fig. \ref{fig:model_pipeline}) predicts an affine transformation matrix from the point itself and its neighbours, and applies it directly to the input point cloud.

\subsubsection{Cross Learning}
The upper branch of the network architecture illustrated in Fig. \ref{fig:model_pipeline}, is the Autoencoder for cross learning between the point cloud from radar and camera domain. The Encoder contains three successive EdgeConv blocks, where the graph is dynamically updated after each one. Afterwards, their multi-scale outputs are concatenated together, followed by one fully-connected layer to form 136 dimensional latent space. Then, the Decoder is used to reconstruct camera skeletons from prior latent feature vector, having similar structure with the Encoder, while only two EdgeConv blocks present. Finally, after going through the last fully-connected layer (51), the point cloud is reshaped to $17 \times 3$, representing 17 joints of the person in $(x,y,z)$ coordinates. The number $k$ for computing $k$-NNs is set as 3 for all EdgeConv blocks. 

\subsubsection{Classification}
After obtaining the latent features from Encoder, the lower branch shown in Fig. \ref{fig:model_pipeline} extracts both spatial and temporal information of each gesture motion sequence with length $l=30$. The latent feature vectors of each frame in the sequence are firstly stacked to create $l \times 136$ tensor, fed as input into two 64-units Bidirectional \ac{lstm} layers \cite{mmpoint_gnn}. The last two fully-connected layer (64,5) produce 5 dimensional feature vector for gesture classification. 
\section{Experimental}
We used a synchronized camera-radar setup as introduced in Section 3 to perform seamless multi-modal data acquisition to train our proposed model and then for evaluation. The training data is balanced and consists of 1773 recordings and 392 recordings (sequence length $l=30$ for each recording) for evaluation. The evaluation dataset doesn't include any camera data as only radar data is used as input during inference.

%%%%%%%Confusion matrix%%%%
\begin{figure}
    \centering
	\includegraphics[width=0.395\textwidth]{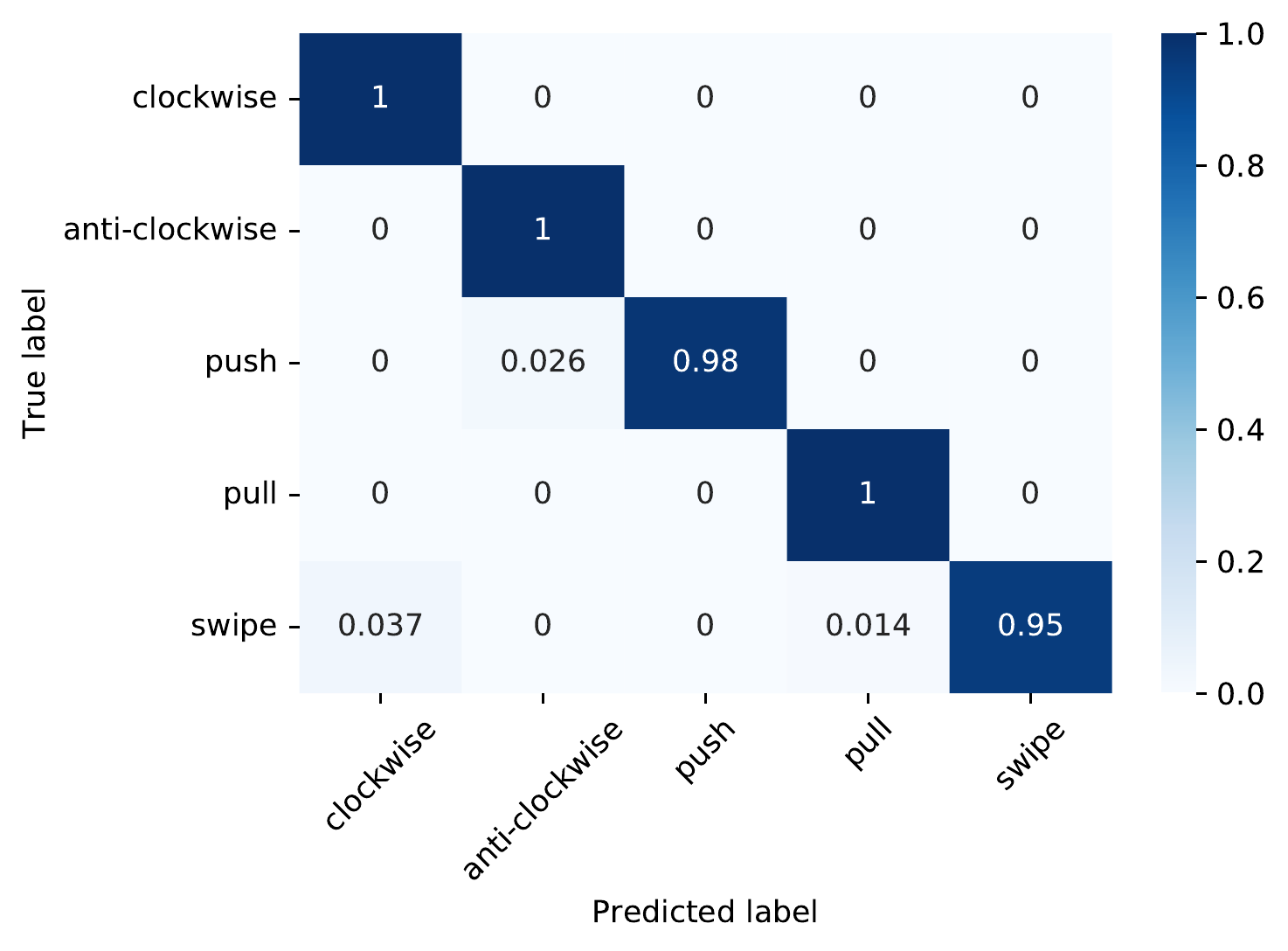}
	\caption{Confusion matrix.}
	\vspace{-4mm}
	\label{fig:confusion_matrix}
\end{figure}

We define five macro-gestures as classes that involve a complete hand movement. The five classes are swipe - where the hand is moved horizontally from one end to another, push- where the hand is moved toward the radar from the body, pull - where the hand is moved from away from the radar towards the body, clockwise - where a big circular movement is done by the hand in a clockwise manner and anti-clockwise - where the same is done but in an anti-clockwise manner. The choice of gestures was made based on their relevance to past literature and simplicity. The training data involved five volunteers who performed the gesture at a distance ranging between 1m -2 m with minimal prior instructions and switched between left and right hands and their distance from the radar. The evaluation dataset was similarly performed by another five volunteers but in a different room and radar orientation setup to demonstrate the proposed architecture's generalization capability. Furthermore, the room setup for the evaluation dataset contained more reflecting objects so as to check our model's ability to suppress such reflections.

%%%%%%%Result%%%%
\begin{figure}
    \centering
	\includegraphics[width=0.395\textwidth]{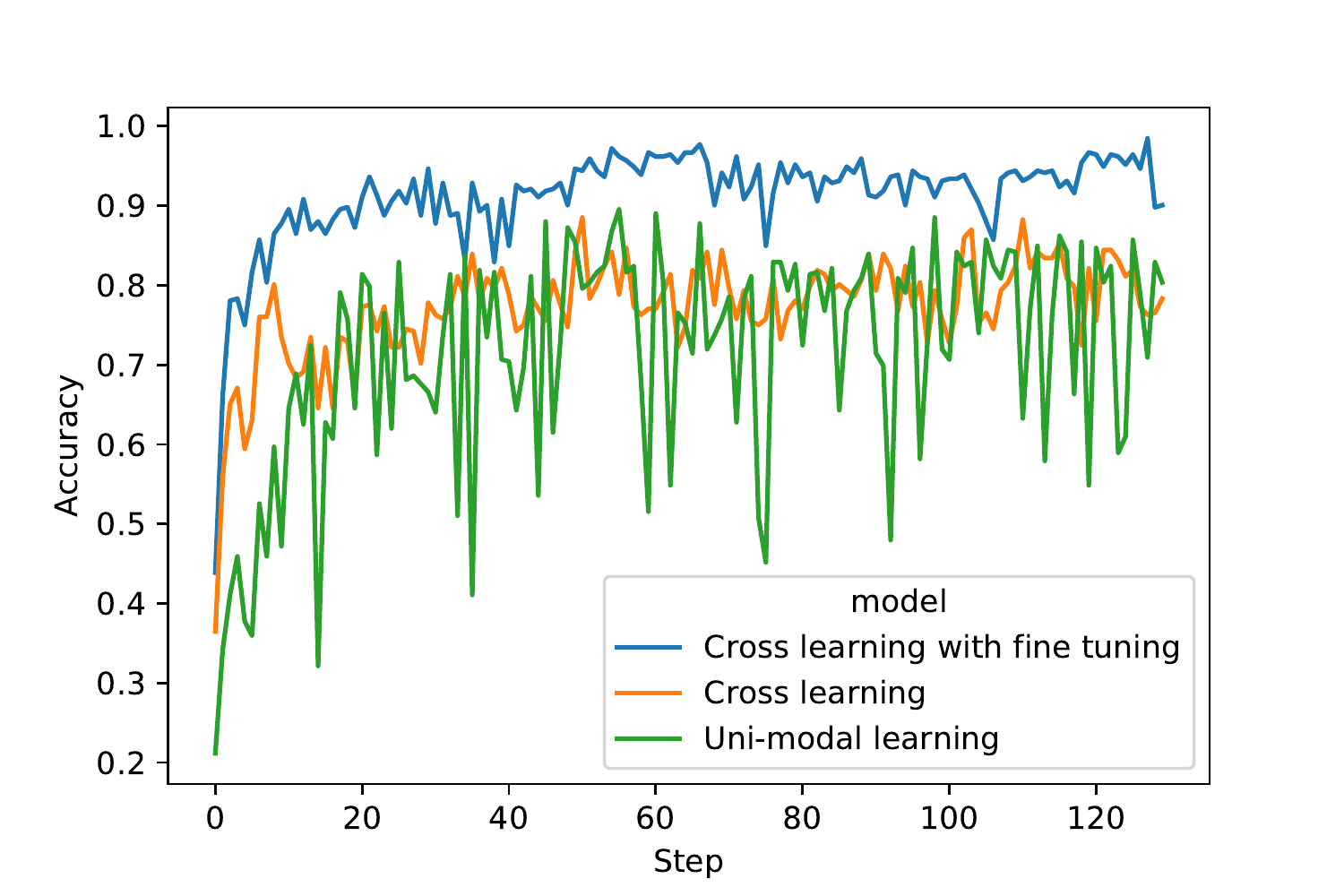}
	\caption{Accuracy comparison among uni-modal, cross learning and cross learning with fine-tuning.}
	\vspace{-4mm}
	\label{fig:result}
\end{figure}

The upper branch in Fig. \ref{fig:model_pipeline} is first trained with synchronized radar point cloud and camera skeletons data using \acrfull{mse} loss, and then we save the \textbf{input transformation} and \textbf{Encoder} modules for the training of classification in the lower branch by Cross-Entropy and Triplet losses combined in the same weight, which can be set as frozen or not for fine-tuning. 
The overall accuracy for the proposed model is 98.4\% compared to from 45\% to 90\% for the baseline model, which is trained in a uni-modal fashion. Fig. \ref{fig:confusion_matrix} depicts the confusion matrix for our proposed model and Fig. \ref{fig:result} represents the training and testing accuracy of different setups. It clearly demonstrates the superiority of the proposed learning approach over a uni-modal approach with better accuracy, robustness, and stable learning.

\section{Conclusion}
The paper proposes a low-cost radar-based gesture recognition system with an overall accuracy of 98\% where the subject performs macro gestures at a varying range between 1m to 2m. We introduce a novel architecture based on \ac{gnn}s that enables knowledge distillation from camera point clouds to radar point clouds, allowing the system to learn better underlying relationships between the radar points and suppress other possible targets in the field of view.

%---------------------------------------------------------------------
% Acknowledgement 
%---------------------------------------------------------------------
%\section*{Acknowledgment}
%    Acknowledgement goes here
%---------------------------------------------------------------------
% References
%---------------------------------------------------------------------
\newpage
\bibliographystyle{IEEEtran}
\bibliography{references}

\begin{thebibliography}{10}
\providecommand{\url}[1]{#1}
\csname url@samestyle\endcsname
\providecommand{\newblock}{\relax}
\providecommand{\bibinfo}[2]{#2}
\providecommand{\BIBentrySTDinterwordspacing}{\spaceskip=0pt\relax}
\providecommand{\BIBentryALTinterwordstretchfactor}{4}
\providecommand{\BIBentryALTinterwordspacing}{\spaceskip=\fontdimen2\font plus
\BIBentryALTinterwordstretchfactor\fontdimen3\font minus
  \fontdimen4\font\relax}
\providecommand{\BIBforeignlanguage}[2]{{%
\expandafter\ifx\csname l@#1\endcsname\relax
\typeout{** WARNING: IEEEtran.bst: No hyphenation pattern has been}%
\typeout{** loaded for the language `#1'. Using the pattern for}%
\typeout{** the default language instead.}%
\else
\language=\csname l@#1\endcsname
\fi
#2}}
\providecommand{\BIBdecl}{\relax}
\BIBdecl

\bibitem{gesture_smart_home_radar}
Q.~Wan, Y.~Li, C.~Li, and R.~Pal, ``Gesture recognition for smart home
  applications using portable radar sensors,'' in \emph{2014 36th Annual
  International Conference of the IEEE Engineering in Medicine and Biology
  Society}, 2014, pp. 6414--6417.

\bibitem{ges_radar_car}
K.~A. Smith, C.~Csech, D.~Murdoch, and G.~Shaker, ``Gesture recognition using
  mm-wave sensor for human-car interface,'' \emph{IEEE Sensors Letters},
  vol.~2, no.~2, pp. 1--4, 2018.

\bibitem{gesture_recog_4154947}
S.~Mitra and T.~Acharya, ``Gesture recognition: A survey,'' \emph{IEEE
  Transactions on Systems, Man, and Cybernetics, Part C (Applications and
  Reviews)}, vol.~37, no.~3, pp. 311--324, 2007.

\bibitem{Hayashi2021}
E.~Hayashi, J.~Lien, N.~Gillian, L.~Giusti, D.~Weber, J.~Yamanaka, L.~Bedal,
  and I.~Poupyrev, ``{RadarNet}: Efficient gesture recognition technique
  utilizing a miniature radar sensor,'' in \emph{Proceedings of the 2021 {CHI}
  Conference on Human Factors in Computing Systems}.\hskip 1em plus 0.5em minus
  0.4em\relax {ACM}, may 2021.

\bibitem{Stephan2021a}
M.~Stephan, S.~Hazra, A.~Santra, R.~Weigel, and G.~Fischer, ``People counting
  solution using an {FMCW} radar with knowledge distillation from camera
  data,'' in \emph{2021 {IEEE} Sensors}.\hskip 1em plus 0.5em minus 0.4em\relax
  {IEEE}, october 2021.

\bibitem{Stephan2022}
M.~Stephan, L.~Servadei, J.~Arjona-Medina, A.~Santra, R.~Wille, and G.~Fischer,
  ``Scene-adaptive radar tracking with deep reinforcement learning,''
  \emph{Machine Learning with Applications}, vol.~8, p. 100284, june 2022.

\bibitem{servadei2021label}
L.~Servadei, H.~Sun, J.~Ott, M.~Stephan, S.~Hazra, T.~Stadelmayer, D.~S.
  Lopera, R.~Wille, and A.~Santra, ``Label-aware ranked loss for robust people
  counting using automotive in-cabin radar,'' \emph{IEEE International
  Conference on Acoustics, Speech and Signal Processing (ICASSP)}, may 2022.

\bibitem{santra2020deep}
\BIBentryALTinterwordspacing
A.~Santra and S.~Hazra, \emph{Deep Learning Applications of Short Range
  Radars}, ser. Artech House radar library.\hskip 1em plus 0.5em minus
  0.4em\relax Artech House, 2020. [Online]. Available:
  \url{https://books.google.de/books?id=Qb-VzQEACAAJ}
\BIBentrySTDinterwordspacing

\bibitem{8542778}
S.~Hazra and A.~Santra, ``Robust gesture recognition using millimetric-wave
  radar system,'' \emph{IEEE Sensors Letters}, vol.~2, no.~4, pp. 1--4, 2018.

\bibitem{8999087}
------, ``Radar gesture recognition system in presence of interference using
  self-attention neural network,'' in \emph{2019 18th IEEE International
  Conference On Machine Learning And Applications (ICMLA)}, 2019, pp.
  1409--1414.

\bibitem{7131232}
P.~Molchanov, S.~Gupta, K.~Kim, and K.~Pulli, ``Short-range fmcw monopulse
  radar for hand-gesture sensing,'' in \emph{2015 IEEE Radar Conference
  (RadarCon)}, 2015, pp. 1491--1496.

\bibitem{wang2019dynamic}
\BIBentryALTinterwordspacing
Y.~Wang, Y.~Sun, Z.~Liu, S.~E. Sarma, M.~M. Bronstein, and J.~M. Solomon,
  ``Dynamic graph cnn for learning on point clouds,'' \emph{ACM Trans. Graph.},
  vol.~38, no.~5, october 2019. [Online]. Available:
  \url{https://doi.org/10.1145/3326362}
\BIBentrySTDinterwordspacing

\bibitem{wu2019detectron2}
Y.~Wu, A.~Kirillov, F.~Massa, W.-Y. Lo, and R.~Girshick, ``Detectron2,''
  \url{https://github.com/facebookresearch/detectron2}, 2019.

\bibitem{cocodataset}
T.-Y. Lin, M.~Maire, S.~Belongie, L.~Bourdev, R.~Girshick, J.~Hays, P.~Perona,
  D.~Ramanan, C.~L. Zitnick, and P.~Dollár, ``Microsoft coco: Common objects
  in context,'' 2015.

\bibitem{personreid}
H.~Luo, Y.~Gu, X.~Liao, S.~Lai, and W.~Jiang, ``Bag of tricks and a strong
  baseline for deep person re-identification,'' in \emph{2019 IEEE/CVF
  Conference on Computer Vision and Pattern Recognition Workshops (CVPRW)},
  2019, pp. 1487--1495.

\bibitem{epipolargeometry}
\BIBentryALTinterwordspacing
Z.~Zhang, \emph{Epipolar Geometry}.\hskip 1em plus 0.5em minus 0.4em\relax
  Boston, MA: Springer US, 2014, pp. 247--258. [Online]. Available:
  \url{https://doi.org/10.1007/978-0-387-31439-6_128}
\BIBentrySTDinterwordspacing

\bibitem{tesla_rapture}
D.~Salami, R.~Hasibi, S.~Palipana, P.~Popovski, T.~Michoel, and S.~Sigg,
  ``Tesla-rapture: A lightweight gesture recognition system from mmwave radar
  sparse point clouds,'' \emph{IEEE Transactions on Mobile Computing}, pp.
  1--1, 2022.

\bibitem{ldc_ges_fmcw}
Z.~Zhang, Z.~Tian, and M.~Zhou, ``Latern: Dynamic continuous hand gesture
  recognition using fmcw radar sensor,'' \emph{IEEE Sensors Journal}, vol.~18,
  no.~8, pp. 3278--3289, 2018.

\bibitem{comprehensive_mpoint}
\BIBentryALTinterwordspacing
G.~Zhang, X.~Geng, and Y.-J. Lin, ``Comprehensive mpoint: A method for 3d point
  cloud generation of human bodies utilizing fmcw mimo mm-wave radar,''
  \emph{Sensors}, vol.~21, no.~19, 2021. [Online]. Available:
  \url{https://www.mdpi.com/1424-8220/21/19/6455}
\BIBentrySTDinterwordspacing

\bibitem{radar_pointgnn}
P.~Svenningsson, F.~Fioranelli, and A.~Yarovoy, ``Radar-pointgnn: Graph based
  object recognition for unstructured radar point-cloud data,'' in \emph{2021
  IEEE Radar Conference (RadarConf21)}, 2021, pp. 1--6.

\bibitem{jin2020mmfall}
F.~Jin, A.~Sengupta, and S.~Cao, ``mmfall: Fall detection using 4-d mmwave
  radar and a hybrid variational rnn autoencoder,'' \emph{IEEE Transactions on
  Automation Science and Engineering}, pp. 1--13, 2020.

\bibitem{3d_radar_pointcloud_gen}
\BIBentryALTinterwordspacing
K.~Qian, Z.~He, and X.~Zhang, ``3d point cloud generation with millimeter-wave
  radar,'' \emph{Proc. ACM Interact. Mob. Wearable Ubiquitous Technol.},
  vol.~4, no.~4, december 2020. [Online]. Available:
  \url{https://doi.org/10.1145/3432221}
\BIBentrySTDinterwordspacing

\bibitem{rodnet}
\BIBentryALTinterwordspacing
Y.~Wang, Z.~Jiang, Y.~Li, J.-N. Hwang, G.~Xing, and H.~Liu, ``{RODNet}: A
  real-time radar object detection network cross-supervised by camera-radar
  fused object 3d localization,'' \emph{{IEEE} Journal of Selected Topics in
  Signal Processing}, vol.~15, no.~4, pp. 954--967, june 2021. [Online].
  Available: \url{https://doi.org/10.1109%2Fjstsp.2021.3058895}
\BIBentrySTDinterwordspacing

\bibitem{9531070}
D.~S. Lopera, L.~Servadei, G.~N. Kiprit, S.~Hazra, R.~Wille, and W.~Ecker, ``A
  survey of graph neural networks for electronic design automation,'' in
  \emph{2021 ACM/IEEE 3rd Workshop on Machine Learning for CAD (MLCAD)}, 2021,
  pp. 1--6.

\bibitem{DBLP:journals/corr/QiSMG16}
R.~Q. Charles, H.~Su, M.~Kaichun, and L.~J. Guibas, ``Pointnet: Deep learning
  on point sets for 3d classification and segmentation,'' in \emph{2017 IEEE
  Conference on Computer Vision and Pattern Recognition (CVPR)}, 2017, pp.
  77--85.

\bibitem{mmpoint_gnn}
P.~Gong, C.~Wang, and L.~Zhang, ``Mmpoint-gnn: Graph neural network with
  dynamic edges for human activity recognition through a millimeter-wave
  radar,'' in \emph{2021 International Joint Conference on Neural Networks
  (IJCNN)}, 2021, pp. 1--7.

\end{thebibliography}

%\printglossaries % prints the acronym list only for debugging- to be deleted
\end{document}